\documentclass{iau_FM}
\usepackage{graphicx}
\usepackage{url} 
\usepackage[colorlinks = true, allcolors = blue]{hyperref} 

\title[Starburst heating in multiphase galactic outflows] 
{Starburst heating and synthetic ion column densities in multiphase galactic outflows}

\author[D. Villarruel, W. Banda-Barrag\'an and B. Casavecchia]   
{Daniel Villarruel$^{1,\dag}$,
Wladimir E. Banda-Barrag\'an$^{1,2}$
\and Benedetta Casavecchia$^2$}

\affiliation{$^1$Escuela de Ciencias F\'isicas y Nanotecnolog\'ia, Universidad Yachay Tech, Hacienda San Jos\'e S/N, 100119 Urcuqu\'i, Ecuador $^{\dag}$Email: {\tt \href{mailto:daniel.villarruel@yachaytech.edu.ec}{daniel.villarruel@yachaytech.edu.ec}} \\[\affilskip]

$^{2}$Hamburger Sternwarte, University of Hamburg, Gojenbergsweg 112, 21029 Hamburg, Germany\\

$^3$Max-Planck-Institut für Astrophysik, Karl-Schwarzschild-Strasse 1, 85748 Garching b. München, Germany}

\pubyear{2024}
\jname{Astronomy in Focus, Volume 1} 
\editors{Worral D., ed.}
\begin{document}

\maketitle

\begin{abstract}


Stellar-driven galactic winds are multiphase outflows of energy and matter connecting the interstellar and circumgalactic media (CGM) with the intergalactic medium. Galactic winds contain a hot and diffuse phase detected in X-rays, and a cold and dense phase detected via emission and absorption lines from the ions populating the outflow. The ion production within galactic winds largely depends on the background UV radiation field produced by star formation, and this in turn depends on the age of the starburst, the gas metallicity, the proximity of the outflowing gas to the central star-forming regions. Our study probes the influence of the proximity of wind-cloud systems to the UV background source, and the effects of magnetic fields on the N\,{\sc v} ion production through the analysis of synthetic column densities and spectral lines. We utilise magnetohydrodynamical simulations to study weakly-magnetised wind-cloud systems, and extract synthetic spectral lines with Trident and yt. Our simulations indicate that magnetic fields transverse to the wind have a shielding effect on dense gas, producing broader N\,{\sc v} absorption lines. Also, a weak (distant) UV background produces N\,{\sc v} only in the outer cloud layers with no spectral signature, while a strong (nearby) UV background produces it in the cloud core with a narrow spectral line. Overall, transverse magnetic fields and a UV radiation at $50\,\rm kpc$ produce the stronger N\,{\sc v} spectral lines.


%
\keywords{Galaxy: evolution, ISM: clouds, MHD, methods: numerical}
\end{abstract}

\firstsection 
\section{Introduction}



The exact mechanisms of galaxy evolution that provide an external supply of gas to the interstellar medium (ISM) in order to maintain star formation and sustain small fractions of baryons and metals are still unknown (\cite{Tumlinson2017}). Galatic outflows fueled by supernovae or active galactic nuclei are essential to regulate the exchange of gases between the ISM and the intergalactic medium (IGM) and provide the observed chemical enrichment in galaxies and low metal retention (\cite{Peeples2014}), The circumgalactic medium (CGM) acts as a transition region where this exchange takes place.  Evidence of these outflows is obtained through absorption spectra given different UV background sources, such as quasars. Looking at the starlight produced by a galaxy is especially useful for spectroscopy of the galactic outflows and inflows of the galaxy (\cite{Tumlinson2017}).


One fundamental problem is the extension of lifetime and stability of clouds to the observed time scales given the presence of neutral atoms and high velocities. The addition of magnetic fields of various strengths to simulations suggest higher consistency among models due to the inhibition of small-scale hydrodynamical instabilities, along with cloud lifetime increase for magnetic fields transverse to the direction of the galactic winds by a draping effect, with generally smoother morphologies (\cite{Cottle2020, Casavecchia24}), as well as general lifetime increase including radiative cooling. Obtaining observables is crucial for agreement comparison of computational and observational results of multiphase gas, including absorption lines (\cite{delaCruz2021, Casavecchia24}) with the aim to improve theoretical understanding. Nevertheless, similarities are still limited by resolution, reproduction of environmental conditions close to those observed in the CGM and the understanding of the effects of UV sources. In this paper we present the effects different UV backgrounds on the chemistry of high-resolution wind-cloud systems affected by magnetic fields in different orientations through the production of synthetic ion column densities using a Python suite developed by our group.

\section{Wind-cloud simulations}

Our 3D simulations of wind-cloud in the CGM are obtained using the PLUTO code (see \cite{Mignone2007}) and the initial conditions provided in \cite{Casavecchia24} for two different orientations of a $0.2$ $\mu$G magnetic field with a $0.3$ pc resolution. One magnetic field (AL model) is aligned to the direction of the wind, while the other one (TR model) is transverse to it. PLUTO numerically solves the equations for the mass, energy and momentum conservation, magnetic induction, solenoidal condition and an additional advection equation, as shown in \cite{Casavecchia24} using a RK3 time-marching algorithm and a Lax–Friedrichs solver (\cite{Mignone2007}). The physical domain comprises dimensions of $120\times 240\times 120$ pc, which correspond to a computational grid resolution of $384\times 768\times 384$ with outflow boundary conditions introduced on all sides.

The wind-cloud models are based on previous models presented in \cite{Banda2016}, and feature clouds with idealised smoothed density distributions, an initial spherical radius of $10$ pc, mass-weighted number density $\langle n_c\rangle=6.9\times 10^{-1}$ cm$^{-3}$, mass-weighted number temperature  $\langle T_c\rangle = 5.5 \times 10^{3}$ K, and subject to winds with Mach number $\mathcal{M}_w=4$ and number density $\hat{n}_w=10^{-3}$ cm$^{-3}$ \cite{Casavecchia24}. These initial conditions are chosen to mimic the expected environmental conditions of the circumgalactic medium (CGM) in the Milky Way (MW, \cite{Richter2017}).

The Cloudy code (\cite{Ferland98}) provides the number density distribution for different ion species in the wind-cloud models by running it with custom UV backgrounds derived from Starburst99 (see \cite{Leitherer99}) models. We consider a metallicity close to solar values and star formation rates similar to the Milky Way, with the wind-cloud system at $50$ kpc from the source ($50$kpc-B). Using the cloud crushing time $t_{cc}$ (\cite{Jones96}) it is possible to reduce the analysis to specific moments of the system evolution. For our analysis we take snapshots at $t/t_{cc}=0.2$, $0.9$, $1.7$ and $2.5$ that correspond to times $t=0.2$, $1.2$, $2.2$ and $3.2$ Myr.

We obtain synthetic observables in the form of column number densities and velocity spectra from the wind-cloud models and the custom UV backgrounds using Python and supporting astrophysical software. The script \texttt{ionbalance.py} uses a SED from Starburst99 models (\cite{Leitherer99}) and provides adequate units and files for further processing with the \texttt{CIAOLoop} tools (\url{https://github.com/brittonsmith/cloudy_cooling_tools}) for Cloudy. The resulting cooling rates for a custom set of ion species in the range of temperatures between $10^{4}$ K and $10^{9}$ K are referred to as ion tables. These ion tables are part of the input of the script \texttt{pluto\_trident.py}. The other inputs are the numerical simulations provided by PLUTO with the conditions described earlier.

In addition, \texttt{pluto\_trident.py} uses TRIDENT (\cite{Hummels2017}) and YT (\cite{Turk2011}) to load the ion tables previously generated with the UV background source at different distances to the wind-cloud system. TRIDENT and YT produce synthetic column density maps and spectra from the number densities of ions derived from the densities, temperatures, metallicity, ionisation fractions and redshift. The script produces text files containing the column densities and velocity spectra for a custom set of ions, which is processed to obtain the figures of those observables.

\section{Results}

\begin{figure}[ht!]
\begin{center}
 \includegraphics[width=5.2in]{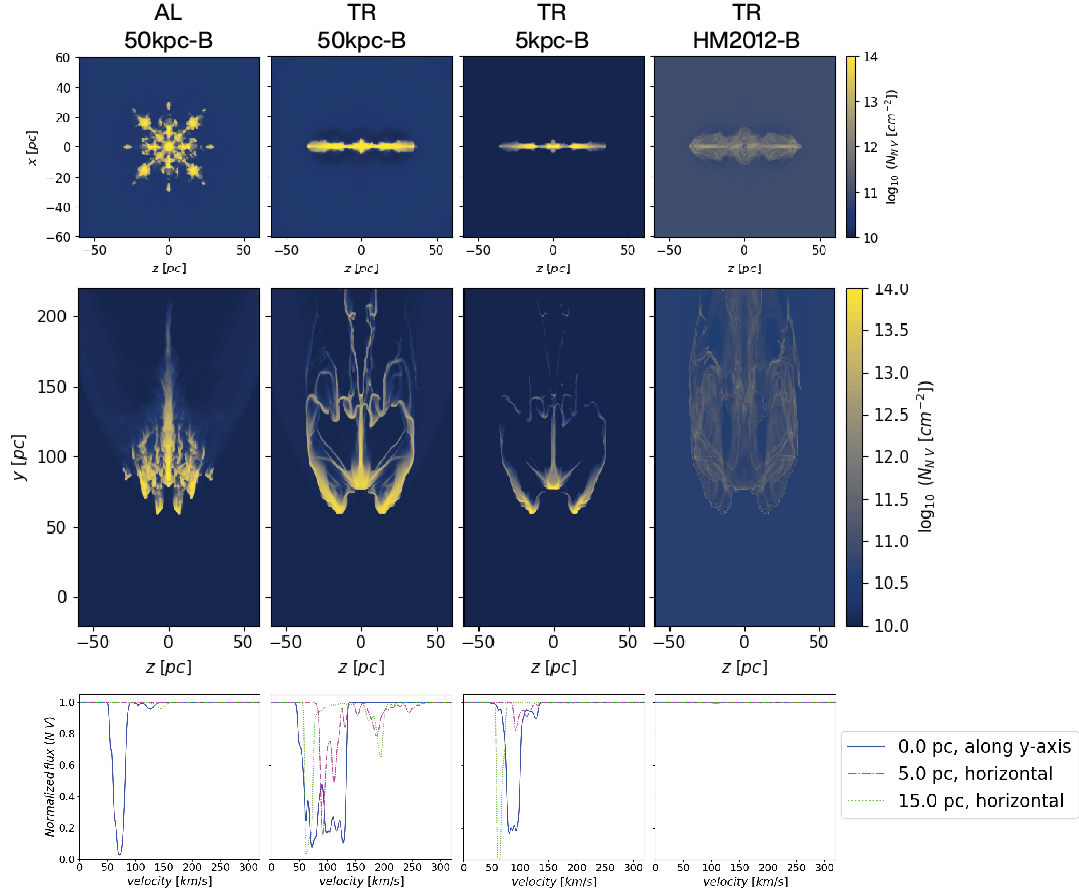} 
 \caption{Synthetic column densities and spectra for N$_V$ for the AL and TR models with different UV backgrounds for down-the-barrel (top) and edge-on (middle) views. Rays for spectral lines pass through the center of the cloud along the y-axis (full line), 5pc from the center in the z-axis (dashed line) and 10pc from the center in the z-axis (dotted line).}
   \label{fig1}
\end{center}
\end{figure}

Figure \ref{fig1} shows the column number densities for N\,{\sc v} for the AL and TR models using Starburst99. The magnetic field has significant effects on the morphology of clouds in both models, shown by the splitting into cloudlets and their different orientations. The down-the-barrel view depicts symmetrical distributions around the $y-$axis for the AL model and an elongated distributed along the $z-$axis in the TR model due to the drapoing effect: field lines wrapped around the cloud in the xy plane creates a net inward force, promoting RT instabilities. The edge-on view for the AL model exhibits a rope-like extended filament, and all the cloud gas is confirmed to be distributed symmetrically around the $y-$axis, while the TR model presents a density distribution limited to the $z-$axis and larger and broader extensions in the $y-$axis.

In addition, we examine synthetic spectral lines (see Fig. \ref{fig1}) for both models. The velocity spectra are calculated with Trident for three different rays passing through the cloud given the observed differences in the density distributions: the first ray passes trough the y-axis, the second ray is parallel to the first but displaced 5 pc in the positive the $z$ direction, and the third is displaced 10 pc also in the positive $z$ direction. All absorption spectra are obtained at $2.2$ Myr. The TR model exhibits deeper and broader absorption lines than the AL model. The transverse magnetic field shows significant lines for the second and third rays due to the distribution of cloud material in the $z-$axis.

In addition, we study the effects of two additional UV backgrounds at different distances from the central starburst on the wind-cloud simulations. The UV source at $5$ kpc ($5$kpc-B) has otherwise the same parameters as the $50$kpc-B. To study the effects of different SEDs, we use the known metagalactic UV background reported by \cite{hm2012} (HM2012-B). We analyse the observables using these UV background sources with special emphasis on the TR model. Morphologically, the fundamental shape of the clouds in the TR model using $5$kpc-B and HM2012-B at $2.2$ Myr is maintained, with differences in density. Features of  N\,{\sc v} in $5$kpc-B are lower and more spatially spread and centred around the edges of the cloud, while for HM2012-B density is extremely low, with regions of high density limited to the edges of the cloud.

We obtained spectral lines using the same procedure described for $50$kpc-B. Using $5$kpc-B N\,{\sc v} lines are noticeably narrower but still present. For HM2012-B absorption lines are not evident. Ionisation in $5$kpc-B produces lower number densities concentrated around the front of the cloud, while for HM2012-B N\,{\sc v} is found in the environment at low number densities, with slightly higher concentrations in the cloud.

\section{Conclusions}

\begin{itemize}
    \item A magnetic field transverse to the direction of the wind produces a shielding effect owing to the draping of field lines around the cloud.
    \item The distance of the wind-cloud system to the UV source plays a crucial role at determining the ion chemistry of the cloud. For weak (distant) sources such as HM2012-B, N\,{\sc v} appears thinly spread at the outer layers of the cloud; for medium-distance sources ($50$kpc-B) the ion density is larger and concentrated around the edges of the clouds, with its velocity spectra spanning larger ranges. The closest source ($5$kpc-B) portrays N\,{\sc v} condensed mainly at the front of the cloud, showing a narrower spectral line.
    \item Using a weak metagalactic UV background is not sufficient for recreating the ISM or the CGM conditions close to the central starburst. In comparison with the other UV backgrounds, N\,{\sc v} is concentrated in the outer parts of the cloud and spectral lines are almost not present.
\end{itemize}

Controlled simulations of small wind-cloud systems are crucial to understanding the microphysics of the CGM around SF galaxies. We have shown that different UV backgrounds influence the resulting spectral lines. We plan extend this work in the future to account for UV heating rates in the PLUTO code, studying additional distances from the central starburst, and improving our Python code.  Our group's efforts aim to produce a more universal and user-friendly programming tool to ease the pipeline of processes needed to extract observables that are intended to be comparable with actual observations.\par

\textbf{Acknowledgements:} The authors gratefully acknowledge the Gauss Centre for Supercomputing e.V. (\url{www.gauss-centre.eu}) for funding this project by providing computing time (via grant pn34qu) on the GCS Supercomputer SuperMUC-NG at the Leibniz Supercomputing Centre (\url{www.lrz.de}). In addition, the authors thank CEDIA (\url{www.cedia.edu.ec}) for providing access to their HPC cluster as well as for their technical support. We also thank the developers of the PLUTO code for making this hydrodynamic code available to the community.

\end{document}